\documentstyle[12pt,epsf]{article}
\begin{document}
 
\newcommand{\be}{\begin{eqnarray}}
\newcommand{\ee}{\end{eqnarray}}
\begin{flushright}
LBNL-52160
\end{flushright}
\vspace{3mm} 
\begin{center}
 
{\Large
{$\Xi^-$ and $\Omega$ Distributions in Hadron-Nucleus Interactions\footnote{
This work was supported in part by the Director, Office of Science,
Division of Nuclear Physics of the Office of High Energy
and Nuclear Physics of the U. S. Department of Energy under Contract
Number DE-AC03-76SF00098.}
}} \\[8ex]

R. Vogt$^{a,b}$ and T.D. Gutierrez$^a$\\[2ex]
 $^a$Physics Department\\
 University of California at Davis\\
 Davis, California\quad 95616  \\
 and\\
 $^b$Nuclear Science Division\\
 Lawrence Berkeley National Laboratory\\
 Berkeley, California\quad 94720 \\[6ex]

{\bf ABSTRACT}
\end{center}
\begin{quote}
Strange baryons have long been known to exhibit a leading particle effect.
A recent comparison of $\Xi^-$ production in $\pi^-$, $n$, and $\Sigma^-$
interactions with nuclei show this effect clearly.  These data are 
supplemented by earlier 
measurements of $\Xi^-$ and $\Omega$ production by a $\Xi^-$ beam.  We
calculate the $\Xi^-$ and $\Omega$ $x_F$ distributions and nuclear dependence
in $hA$ interactions using the intrinsic model.
\end{quote}
\vspace{2mm}
\begin{center}
PACS numbers: 12.38.Lg, 13.85.Ni, 14.20.Jn
\end{center}
\newpage

\section{Introduction}

Leading particle effects, flavor correlations
between the final-state hadron and the projectile valence quarks, have 
long been observed in strange particle production.  Although many experiments
have recently focused on leading charm production 
\cite{E791,ddata1,ddata2,ddata3,ddata4,ddata5,wa891,wa892,SELEX1,SELEX2},
the first data involved strange particles 
\cite{bourq1,bourq2,bourq3,biagi,schn,beret}.  With new data from the WA89
collaboration on $\Xi^-$ production by $\pi^-$, $n$, and $\Sigma^-$ 
projectiles on nuclear targets \cite{wa89xi}, 
in addition to $\Xi^-$ production data from 
$\Xi^-$ beams \cite{biagi}, doubly strange hadron production can be studied
as a function of the number of strange valence quarks in the 
projectile.  We compare our model calculations to both the $x_F$
distributions and the integrated $A$ dependence reported by WA89 \cite{wa89xi}.
We also discuss $\Xi^-$ and $\Omega$ production by the $\Xi^-$ beam
\cite{biagi}. 

The WA89 collaboration used carbon, C, and copper, Cu, 
targets to study the $A$ 
dependence of $\Xi^- (dss)$ production by $\pi^-(\overline u d)$, $n(udd)$, 
and $\Sigma^- (dds)$ beams \cite{wa89xi}.  
The negative beams, $\pi^-$ and $\Sigma^-$, had
an average momentum of 345 GeV with a 9\% momentum spread.  
The neutron beam had a lower momentum with a larger
spread than the negative
beams---the average momentum was 260 GeV with a
15\% variation.  The detected $\Xi^-$ was in the forward $x_F$ 
region, $x_F \geq 0.05$, with low transverse momentum, $p_T \leq 2.5$ GeV.
The data were parameterized in the form
\be
\frac{d\sigma}{dp_T^2 dx_F} \propto (1-x_F)^a e^{-bp_T^2} \, \, .
\label{param1}
\ee
The pion and neutron results agree with 
the functional form of Eq.~(\ref{param1}) over
all $x_F$.  For the pion, $a = 3.8 \pm 0.3$ for C and $4.1 \pm 0.3$ for Cu
while for the neutron $a = 5.0 \pm 0.3$ for C and $4.8 \pm 0.3$ for Cu.  These
results are consistent with expectations from spectator 
counting rules \cite{Gun}, $d\sigma/dx_F \propto (1-x_F)^{2n_s - 1}$.
With an incident gluon, $n_s = 2$ for pions and 3 for neutrons,
consistent with no leading particle effect for projectiles with zero 
strangeness.  There is no significant $A$ dependence of the exponent $a$.

On the other hand, the $\Sigma^-$ data cannot be fit to Eq.~(\ref{param1}) 
for $x_F < 0.4$.  In
the large $x_F$ region, $a = 2.08 \pm 0.04$ for C and $1.97 \pm 0.04$ for Cu.
These results indicate a very hard $x_F$ distribution, 
inconsistent with the counting rules even for a valence quark since $n_s = 2$ 
gives $(1-x_F)^3$.  In addition, at $x_F < 0.4$, the
distribution is independent of $x_F$ for both targets.  Thus
these data show a strong leading particle effect since the $\Xi^-$ has two 
valence quarks in common with the $\Sigma^-$.
The statistics are also sufficient for an observable $A$ dependence 
in the fitted values of $a$.

The integrated $A$ dependence was also reported by WA89
\cite{wa89xi}.  The $A$ dependence of the total cross section
is often parameterized as
\be
\sigma_{pA} = \sigma_{pp} A^\alpha \, \, .
\label{alfdef}
\ee
The integrated $\alpha$ for $\Sigma^-$ production of $\Xi^-$, 
$\alpha = 0.679 \pm 0.011$ \cite{wa89xi}, is 
in relatively good agreement with previous fits.  However, the pion
and neutron data show a closer-to-linear $A$ dependence, 
$\alpha = 0.891 \pm 0.034$ 
and $0.931 \pm 0.046$ respectively.  WA89 attributes this difference to
the fact that two $s \overline s$ pairs must be produced to make the 
final-state $\Xi^-$ and $s \overline s$ pair production would be suppressed 
relative to light $q \overline q$ production.  

WA89 has also measured the dependence of $\alpha$ on $x_F$. 
This dependence, $\alpha(x_F)$, was previously reported for a wide range of 
hadron projectiles \cite{Geist}.  For non-strange hadrons and hadrons with a
single strange quark, there is a common trend with $x_F$.
At $x_F = 0$, $\alpha \approx 0.8$ and decreases to
$\approx 0.5$ at large $x_F$, an overall decrease of $\sim A^{1/3}$ for
$0 < x_F < 1$.  The $\Xi^0$, the only doubly-strange hadron included in
Ref.~\cite{Geist}, is an exception.  In
$pA$ interactions, the $\Xi^0$ has a larger value of $\alpha$ at low $x_F$
\cite{beret}.  A similar effect is observed for $\Xi^-$ production
by WA89.  Their measurements of $\alpha(x_F)$ for $\Xi^-$ from 
pion and neutron beams show that $\alpha \sim 1$ for $x_F
\sim 0.05$, decreasing to $\alpha \sim 0.7$ at higher $x_F$.  
Thus the decrease of $\alpha$ with $x_F$ is also $A^{1/3}$ in this case
although the actual values of $\alpha$ are larger than those for lighter
hadrons \cite{Geist}.  However, for $\Sigma^-$-induced $\Xi^-$ production, 
$\alpha \sim 0.7$ almost independent of $x_F$.  

The other data we consider are $\Xi^- \, {\rm Be} \rightarrow \Xi^-,
\Omega(sss)$ at 116 GeV, measured by Biagi {\it et al.}\ \cite{biagi}.  
In this case, the final-state $\Xi^-$ $x_F$ distribution 
increases with $x_F$, as does the $\Omega$ $x_F$ 
distribution.  This increase could be due in part to
the use of an invariant parameterization\footnotetext{For the two
parameterizations to be equivalent, the right-hand side of Eq.~(\ref{param1})
should be multiplied by $2E/\sqrt{s}$ to obtain the invariant cross section.}
\cite{biagi},
\be
E \frac{d\sigma}{dp^3} \propto (1-x_F)^{a'} e^{-b' p_T^2} \, \, ,
\label{param2}
\ee
which fits the $\Xi^-$ data at $x_F > 0.5$ but only 
approximately fits the $\Omega$ data in this limited region.
The exponent $a'$ was fit in two $p_T^2$ intervals, $p_T^2 < 0.4$ GeV$^2$ and
$0.4 < p_T^2 < 2.9$ GeV$^2$, yielding $a' = -0.45 \pm 0.02$ and 
$-0.18 \pm 0.03$ respectively.  Between the most central measurement, 
$x_F = 0.15$, and the 
projectile fragmentation region, $x_F = 0.85$, the $\Xi^-$ cross section 
increases by a factor of $\sim 40$ in the low $p_T^2$ interval.  

A comparison of these results with incident proton data \cite{biagi}, 
$pA \rightarrow \Xi^- X$ \cite{bourq1,Cardello}
and $pA \rightarrow pX$ \cite{Barton},
showed that, at low $x_F$, the $\Xi^-$ production cross section
is essentially independent of the projectile while, at high $x_F$, the 
$\Xi^-$ and $p$ scattering cross sections are similar.  This behavior
supports valence quark domination at high $x_F$.  The structure of the $\Omega$
$x_F$ distribution is similar:  it is of the same order of magnitude as $pA
\rightarrow \Omega X$ \cite{bourq1} at low $x_F$ 
but is similar to singly strange baryon production 
by protons, $pA \rightarrow \Lambda X$ and $\Sigma^+ X$, 
\cite{bourq1,Cardello,Skubic} at high $x_F$.

Since only one target was used, $\alpha = 0.6$ was assumed in
Eq.~(\ref{alfdef}) to obtain 
the per nucleon cross sections.  This extrapolated
cross section is a factor of $1.5-2$ higher than those on hydrogen
targets \cite{biagi}.  An extrapolation with $\alpha = 1$ 
gives better agreement with the hydrogen target data, at least for
$\Xi^-$ production.

We employ the intrinsic model \cite{intc1,intc2,VB,VBlam,GutVogt1}
as developed for strangeness production in Ref.~\cite{GutVogtIS}. 
In the intrinsic model, a hadron can fluctuate into Fock
state configurations with a combination of light and strange quark pairs. 
The heavier quarks in the configuration are comoving with the other
partons in the Fock state and thus can coalesce with these comoving partons
to produce strange hadrons at large $x_F$. 
The model combines leading-twist production of $s \overline s$ pairs with
intrinsic Fock states with up to nine particles.  Thus coalescence production
of the $\Omega$ from a proton is possible.  

\section{Leading-Twist Production}

We calculate leading-twist strangeness in perturbative QCD, assuming the
strange quark is massive.  When the projectile has nonzero strangeness, we also
consider the possibility of flavor excitation.  We choose proton parton
distribution functions with the lowest possible initial scale $\mu_0^2$ so that
$m_s^2 > \mu_0^2$.  Therefore the baryon parton distribution functions are
based on the GRV 94 LO proton parton distributions \cite{GRV94}.
We use the most recent pion parton densities by Gl\"{u}ck, Reya
and Schienbein \cite{GRSpi2}.  
To be conservative, we assume that the scale $\mu$ at which the strong coupling
constant and the parton densities are evaluated is $\mu = 2m_T$ where $m_T = 
\sqrt{p_T^2 + m_s^2}$ and $m_s = 500$ MeV.
The $x_F$ distributions, obtained by integrating Eq.~(\ref{param1}) or
(\ref{param2}) over $p_T$, are dominated by low $p_T$ production.

We treat the strange quark as heavy, as in Ref.~\cite{GutVogtIS}, 
rather than as a massless parton in jet-like 
processes.  Treating the strange quark as a jet means 
selecting a minimum $p_T$ to keep the cross section finite.  A large minimum
$p_T$ compatible with hard scattering is incompatible with the
assumption of intrinsic production, inherently a low $p_T$
process \cite{VB}.  Strange hadrons can either be produced directly in jet
production or by the fragmentation of light quark
and gluon jets.  However, there is no indication that these data originate 
from jets.

The leading-twist $x_F$ distribution of heavy quark production \cite{VBH2} 
is denoted by $F$,
\be
F \equiv
\frac{d\sigma_{\rm ltf}^S}{dx_F} = 
\frac{\sqrt{s}}{2} \int dz_3\, dy_2\, dp_T^2  x_a x_b H_{AB}(x_a,x_b,\mu^2)
\frac{1}{E_1}\ \frac{D_{S/s}(z_3)}{z_3}\ \, \, ,
\label{ltfus}
\ee
where $A$ and $B$ are the initial hadrons, $a$ and $b$ 
are the interacting partons, 1 and 2 are the 
produced strange quarks and 3 is the final-state strange hadron $S$.
The $x_F$ of the detected quark is $x_F = 2m_T \sinh y/\sqrt{s}$ where $y$
is the rapidity of the quark and $\sqrt{s}$ is the hadron-hadron center of mass
energy.  We assume the simplest possible fragmentation function,
\be D_{S/s}(z) = B_S \delta(1-z) \,\, , \label{fusfrag} \ee
with $B_S = 0.1$, assuming that all
10 ground-state strange hadrons are 
produced at the same rate to leading twist \cite{GutVogtIS}.  This choice
of $D_{S/s}$ gives the hardest possible leading twist $x_F$ distribution.

The convolution of the subprocess $q \overline q$
annihilation and gluon fusion cross sections 
with the parton densities is included in $H_{AB} (x_a, x_b, \mu^2)$,
\be
H_{AB}(x_a,x_b,\mu^2) & = & \sum_q [f_q^A(x_a,\mu^2) 
f_{\overline q}^B(x_b,\mu^2) + f_{\overline q}^A(x_a,\mu^2) f_q^B(x_b,\mu^2)] 
\frac{d \widehat{\sigma}_{q \overline q}}{d \hat{t}} \label{hab} \\
 &   & \mbox{} + f_g^A(x_a,\mu^2) f_g^B(x_b,\mu^2) \frac{d
\widehat{\sigma}_{gg}}{d \hat{t}} \, \, , \nonumber
\ee
where $q = u,$ $d$, and
$s$.  Although including the $s$ quark in the sum over $q$ in Eq.~(\ref{hab}) 
could lead to some over counting,
the strange quark contribution to $F$ from non-strange projectiles is 
negligible, less than $0.01$\% for neutron and pion projectiles.  It is
somewhat larger for strange
projectiles, 2.5\% for the $\Sigma^-$ and 5.6\% for the $\Xi^-$ but it is
only significant at large $x_F$.  

Hyperon parton distributions can be
inferred from the proton distributions \cite{GutVogt1} by simple counting
rules.  We can relate the valence $s$ distribution of the $\Sigma^-$,
$f_{s_v}^{\Sigma^-}$, to the proton valence $d$ distribution, $f_{d_v}^p$,
and the valence $d$ distribution in the $\Sigma^-$,
$f_{d_v}^{\Sigma^-}$, 
to the valence $u$ in the proton, $f_{u_v}^p$, so that
\begin{eqnarray}
\int_0^1 dx \, f_{s_v}^{\Sigma^-} (x,\mu^2) & = & \int_0^1 dx \, f_{d_v}^p 
(x,\mu^2) = 1 \, \, , \\
\int_0^1 dx \, f_{d_v}^{\Sigma^-} (x,\mu^2) & = & \int_0^1 dx \, f_{u_v}^p 
(x,\mu^2) = 2 \, \, .
\label{partsumsig}
\end{eqnarray}
We also identify the up quark in the 
sea of the $\Sigma^-$ with the strange quark in the proton sea, 
$f_u^{\Sigma^-}(x,\mu^2) = f_s^p(x,\mu^2)$.  
Similar relations hold for the antiquark distributions.  Likewise,
for the $\Xi^-$, we relate the valence $s$, $f_{s_v}^{\Xi^-}$, to the 
valence $u$ in the proton, $f_{u_v}^p$, and
equate the valence $d$ distributions in both baryons so that,
\begin{eqnarray}
\int_0^1 dx \, f_{s_v}^{\Xi^-} (x,\mu^2) & = & \int_0^1 dx \, f_{u_v}^p 
(x,\mu^2) = 2 \, \, , \\
\int_0^1 dx \, f_{d_v}^{\Xi^-} (x,\mu^2) & = & \int_0^1 dx \, f_{d_v}^p 
(x,\mu^2) = 1 \, \, .
\label{partsumxi}
\end{eqnarray}
Here also, $f_u^{\Xi^-}(x,\mu^2) = f_s^p(x,\mu^2)$.  
The gluon distributions are assumed to
be identical for all baryons, $f_g^p = f_g^{\Sigma^-} = f_g^{\Xi^-}$.  
The leading order
subprocess cross sections for heavy quark production
can be found in Ref.~\cite{Ellis}.  
The fractional momenta carried by the projectile and target partons, 
$x_a$ and $x_b$,
are $x_a = (m_T/\sqrt{s}) (e^y + e^{y_2})$ and $x_b = (m_T/\sqrt{s}) (e^{-y} 
+ e^{-y_2})$.

We have assumed only $gg$ and $q \overline q$ contributions
to massive strange quark production.
We have also checked how the $x_F$ distribution would
change if the strange quark was treated as massless and all $2 \rightarrow 2$
scattering channels were included.  
Jet production of $s$ quarks is through processes such as
$g s \rightarrow g s$, $q s \rightarrow qs$ and $\overline q s \rightarrow 
\overline q s$.  (Similarly for the $\overline s$.)
Including these processes increases the cross section by a factor 
of $4-8$.  While this factor
is not constant, it increases rather slowly with $x_F$ so that the difference
in shape is only important in the region where intrinsic production dominates,
as discussed later.

Contributions from massless $2 \rightarrow 2$ scattering 
increase more rapidly at $x_F >0$ for strange projectiles
because the contribution from, for example, $f_s^{\Sigma^-}(x_a) f_g^p(x_b)$, 
dominates the scattering cross section.  In the infinite momentum frame,
$f_{s_v}^{\Sigma^-} = f_{d_v}^p$, see Eq.~(\ref{partsumsig}), and
$f_{s_v}^{\Sigma^-}$ is large at large $x_a$ while $f_g^p$ increases as $x_b$ 
decreases. To take this into account quantitatively, 
we have incorporated ``flavor excitation'' of massive strange
valence quarks.  The excitation matrix elements for massive quarks
are found in Ref.~\cite{comb}.  The
flavor excitation cross section has a pole when $p_T \rightarrow 0$ so that
a cutoff, $p_{T_{\rm min}}$, is required to keep this cross
section finite, as in jet production.  We employ $p_{T_{\rm min}}
= 2m_s = 1$ GeV.  The leading-twist fusion cross section for strange
projectiles is then augmented by 
\be
X_{p_{T_{\rm min}}} \equiv
\frac{d\sigma_{\rm lte}^S}{dx_F} = 
\frac{\sqrt{s}}{2} \int  dz_3\, dy_2\, dp_T^2 
x_a' x_b' H_{AB}'(x_a',x_b',\mu^2)
\frac{1}{E_1}\ \frac{D_{S/s}(z_3)}{z_3}\ \, \, 
\label{ltexc}
\ee
where
\begin{eqnarray}
H_{AB}'(x_a',x_b',\mu^2) =  f_{s_v}^A(x_a',\mu^2) \{ \sum_q 
[f_q^B(x_b',\mu^2) + f_{\overline q}^B(x_b',\mu^2)] \frac{d
\widehat{\sigma}_{sq}}{d \hat{t}} + f_g^B(x_b',\mu^2) \frac{d
\widehat{\sigma}_{sg}}{d \hat{t}} \} \, \, , \label{habexc}
\end{eqnarray}
$x_a' = (m_Te^y + p_Te^{y_2})/\sqrt{s}$ and $x_b' = (m_Te^{-y} 
+ p_Te^{-y_2})/\sqrt{s}$.  Note that there is no overlap between the processes
included in Eqs.~(\ref{habexc}) and (\ref{hab}) and thus no double
counting. This excitation mechanism is effective
only for hadrons with a strange quark in the final state and thus does not
affect the distributions with a produced $\overline s$.

To summarize, for strange and antistrange final states produced by non-strange 
hadrons,
\be \frac{d\sigma^S_{\rm lt}}{dx_F} =  \frac{d\sigma^S_{\rm ltf}}{dx_F} \equiv
F \, \, ,
\label{fdef}
\ee
as in Eq.~(\ref{ltfus}).  This relation also holds for antistrange final states
from strange hadrons.  However, for strange hadron production by 
hadrons with nonzero strangeness, we also consider
\be \frac{d\sigma^S_{\rm lt}}{dx_F} = \frac{d\sigma^S_{\rm ltf}}{dx_F} 
+ \frac{d\sigma^S_{\rm lte}}{dx_F} \equiv
F + X_{p_{T_{\rm min}}} \, \, 
\label{efdef}
\ee
where flavor excitation, Eq.~(\ref{ltexc}), may play a role.

We remark that the role of flavor excitation in heavy quark production, as
outlined in Ref.~\cite{comb}, is questionable.  It was first proposed as a
leading order contribution to the total cross section and, as such, could be
rather large if the heavy quark distribution in the proton is significant.
However, the proton heavy quark distribution is only nonzero above the
threshold $m_Q$.  In addition, parton distribution functions are defined in the
infinite momentum frame where the partons are treated as massless.  Later
studies at next-to-leading order (NLO) \cite{NDE} showed that these excitation
diagrams are a subset of the NLO cross section and suppressed relative to
fusion production.  They are only a small fraction of the heavy flavor
cross section and thus play no significant role at low energies.  
Strange hadron production at large $x_F$ is then an important test of the
excitation process.  

\begin{figure}[htpb]
\setlength{\epsfxsize=0.95\textwidth}
\setlength{\epsfysize=0.5\textheight}
\centerline{\epsffile{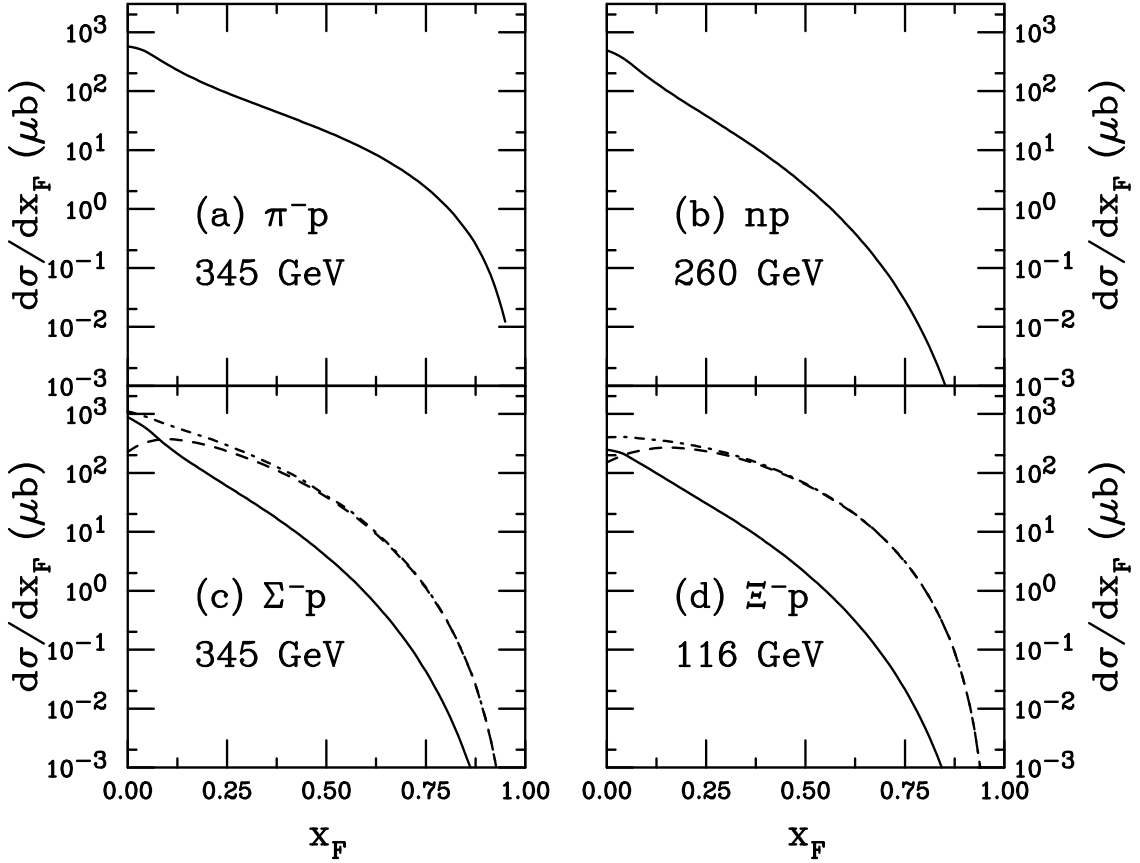}}
\caption[]{Leading-twist strange quark production in
(a) $\pi^- p$ interactions at 345 GeV, (b) $np$ interactions at 260 GeV,
(c) $\Sigma^- p$ interactions at 345 GeV, and (d) $\Xi^- p$ interactions at
116 GeV.  The solid curves are the fusion results, $F$.  For projectiles with
valence strange quarks, the excitation contributions, $X_{p_{T_{\rm min}}}$ 
with $p_{T_{\rm min}} = 1$ GeV, are shown in
the dashed curves. The dot-dashed curves are the total, $F + 
X_{p_{T_{\rm min}}}$.}
\label{fus}
\end{figure}

The leading-twist $x_F$ distributions with all four projectiles are shown in
Fig.~\ref{fus}.  We compare the fusion $x_F$ distributions, $F$, 
in $\pi^- p$ and $\Sigma^- p$ interactions at 345 GeV, $np$ interactions at 
260 GeV, and $\Xi^- p$ interactions at 116 GeV, corresponding to the energies 
we investigate.  The $\pi^- p$ distribution is the broadest because
the $f_{\overline u_v}^{\pi^-}(x_a) f_{u_v}^p(x_b)$ 
contribution hardens the $x_F$
distribution at large $x_F$ where the $q \overline q$ channel dominates.  
When excitation is considered, as in $\Sigma^- p$ and $\Xi^- p$
interactions, the $x_F$ distribution is hardened, particularly through the
$f_s^{\Sigma^-}(x_a') f_g^p(x_b')$ and $f_s^{\Xi^-}(x_a') f_g^p(x_b')$
contributions.  These dominate at large $x_F$ where $f_s^S(x_a')$
is large for valence strange quarks and $f_g^p(x_b')$ is large at small
$x_b'$.  The effect is even stronger for the $\Xi^-$ since it 
has two valence strange quarks.   Note that,
except at small $x_F$, the total leading-twist cross section is equivalent to
$X_{p_{T_{\rm min}}}$.

We can obtain an approximate estimate of the exponent $a$ from
Eq.~(\ref{param1}) from the average $x_F$, $\langle x_F \rangle$,
where 
\be
a = \frac{1}{\langle x_F \rangle} - 2 \, \, .
\label{avea}
\ee
When averaged over $x_F>0$, the values of $a$ obtained are larger than
those measured by WA89, as expected.  For the pion and neutron beams, $a = 5.2$
and 8.7 respectively.  Strangeness production by strange hadrons including
fusion alone also gives large values of $a$, 9 for the $\Sigma^-$ and 7.4 for
the $\Xi^-$.  The $x_F$ distribution of strange quarks produced by flavor
excitation is considerably harder, $a = 3.2$ for the $\Sigma^-$ and 2.2 for
the $\Xi^-$.  Combining the two contributions, as in the dot-dashed curves
in Fig.~\ref{fus}(c) and (d), gives a somewhat larger value of $a$ than for
flavor excitation alone, $a = 4.7$ and 2.9 for $\Sigma^-$ and $\Xi^-$ beams
respectively.  The values of $a$ obtained from Eq.~(\ref{avea}) are all much
higher than those obtained from the data.  Thus the leading twist results alone
cannot explain the shape of the measured $\Xi^-$ $x_F$ distributions.

\section{The Intrinsic Model for Strangeness}

We now briefly discuss the intrinsic model for strangeness production, 
described in detail for $\pi^- p$ interactions in Ref.~\cite{GutVogtIS}.  
Since all the data is at $x_F > 0$, 
we only discuss intrinsic production in the projectile.

The hadron wavefunction is a superposition of Fock state fluctuations
in which the hadron contains one or more ``intrinsic'' $Q \overline Q$ pairs.
These pairs can hadronize when the hadron interacts, breaking the
coherence of the state.  The model, first developed for charm 
\cite{intc1,intc2}, gives heavy quarks a
larger fraction of the projectile momentum due to their greater mass.  
The strange quark is lighter so that the momentum gained is not as large.
However, the intrinsic strangeness probability is larger, 
$P_{\rm is}^5 \sim 2$\%.  For simplicity, we assume that the intrinsic 
probabilities are independent of the valence quark content of the projectile.
Then $P^5_{\rm is}$ is identical for nucleons and hyperons.  The Fock state
probabilities for up to 3$Q \overline Q$ pairs where at least one $Q \overline
Q$ pair is strange are given in
Ref.~\cite{GutVogtIS}.  

In this paper, we focus on $\Xi^-(dss)$ and $\Omega(sss)$ production from
$\pi^- (\overline u d)$, $n(udd)$, $\Sigma^-(dds)$ and $\Xi^-(dss)$
projectiles.  The produced $\Xi^-$ shares one or more valence
quarks with the projectile.  We study $\Omega$ production only by $\Xi^-$
projectiles, with two valence quarks in common.

Once the coherence of the Fock state is broken, the partons in the state can
hadronize in two ways.  The first, uncorrelated fragmentation of the strange
quark, is the same basic mechanism as at leading twist.  However, when the Fock
state fluctuation includes all the valence quarks of the final-state hadron,
these quarks, in close spatial proximity, can coalesce into the final-state
hadron and come on shell.  Thus, to calculate the full strange and 
antistrange hadron $x_F$ distributions in
the intrinsic model, we include uncorrelated fragmentation of the strange
quark in every state considered and coalescence from those states where it is 
possible. In Ref.~\cite{GutVogtIS}, a comparison of the model with strange 
baryon asymmetries suggested that fragmentation may not be an effective
mechanism because when the Fock state has minimal invariant mass, 
fragmentation may cost too much energy.  This conclusion
needs to be checked against inclusive $x_F$ distributions over a broader $x_F$
range.    

In principle, the parton distributions of the hadron can be defined through
such a Fock-state expansion \cite{hoyerroy}.  In each fluctuation, only the
mass distinguishes the light and heavy quark distributions.  Thus it is not
really possible in a given state to separate the ``valence'' and ``sea''
distributions.  All are similar as long as the quarks are light. One
distinguishing feature is our assumption that only strange quarks can produce
strange final-state hadrons by uncorrelated fragmentation.
Thus with hyperon projectiles, uncorrelated
fragmentation may also be possible from Fock
states with only light $Q \overline Q$ pairs.  These intrinsic 
light quark states must be included in the total probability,
as described in Ref.~\cite{GutVogtIS}.  The probabilities
for these states must also be defined.  We assume
\be 
P^5_{\rm iq} & = & \left( \frac{\widehat{m}_s}{\widehat{m}_q} \right)^2 
P^5_{\rm is} \approx 5\% \, \, , \label{piq} \\
P^7_{\rm iqq} & = & \left( \frac{\widehat{m}_s}{\widehat{m}_q} \right)^2
P^7_{\rm isq} = 1.75 \, P^5_{\rm is} \, \, , \label{piqq} \\
P^9_{\rm iqqq} & = & \left( \frac{\widehat{m}_s}{\widehat{m}_q} \right)^4
P^9_{\rm issq} = 1.25 \, P^5_{\rm is} \, \, . \label{piqqq} 
\ee
We further assume that the probabilities for the meson Fock 
configurations are equal to the baryon probabilities.

We have only taken the 10 strange ground state hadrons and 
antihadrons into account.  
We assume that each hadron has a 10\% production probability 
from fragmentation, neglecting the particle masses. The final-state $x_F$
distribution is then equivalent to that of the $s$ or $\overline s$ quark. 
For coalescence, we count the number of 
possible strange and antistrange hadron combinations that can be obtained 
from a given Fock state and
assign each strange hadron or antihadron that fraction of the total.
The possible number of strange hadrons is greater than
the number of possible strange antihadrons.
We clearly err in the overall normalization by simply 
including the ground state strange particles.  However,
the higher-lying resonances have the same quark
content with the same fragmentation
and coalescence distributions since all properties of
the final-state hadrons except their quark content are neglected.

To obtain the total probability of each strange hadron to be produced from
projectile hadron, $h$, in the
intrinsic model, we sum 
the probabilities over all the states with up to 3 intrinsic $Q \overline Q$
pairs.  Thus
\be \frac{dP_S^h}{dx_F} = \sum_n \sum_{m_u} \sum_{m_d} \sum_{m_s} \beta
\left( \frac{1}{10} \frac{dP_{{\rm i} (m_s {\rm s}) (m_u {\rm u}) (m_d {\rm d})
}^{nF}}{dx_F} + \xi \frac{dP_{{\rm i} (m_s {\rm s}) (m_u {\rm u}) (m_d {\rm d})
}^{nC}}{dx_F} \right) \, \, .
\label{intsum}
\ee
To conserve probability, $\beta = 1$ when the hadron is only 
produced by uncorrelated fragmentation and 0.5 when both fragmentation and 
coalescence are possible.  When we assume coalescence production only, we 
set $P^{nF} \equiv 0$ and $\beta \equiv 1$.
The weight of each state produced by coalescence is 
$\xi$ where $\xi = 0$ when $S$ is not produced by coalescence in state
$|n_v  m_s(s \overline s) m_u(u \overline u) m_d(d \overline d) \rangle$.  
The number of up, down and strange $Q \overline Q$ pairs 
is indicated by $m_u$, $m_d$ and $m_s$ respectively.  The total, $m_u + m_d
+ m_s = m$, is defined as $m = (n - n_v)/2$ because each $Q$ in an $n$-parton
state is accompanied by a $\overline Q$. For baryon projectiles, $n =5$, 7, 
and 9 while for mesons $n=4$, 6, and 8.  Depending on the value of $n$, $m_i$
can be 0, 1, 2 or 3, {\it e.g.}\ in a $|uud s \overline s d \overline d d 
\overline d \rangle$ state, $m_u = 0$, $m_d = 2$ and $m_s = 1$ with $m = 3$.  
Note that $m_s=0$ is only possible when $h$ is strange since
no additional $s \overline s$ pairs are thus needed to produce some strange
hadrons by coalescence.
The total probability distributions, $dP_S^h/dx_F$, for $S = \Xi^-$ and 
$\Omega$ are given in the Appendix.

\section{$A$ Dependence of Combined Model}

The total $x_F$ distribution for final-state strange hadron $S$ is the sum
of the leading-twist fusion and intrinsic strangeness components,
\be
\frac{d\sigma^S_{hN}}{dx_F} = \frac{d\sigma^S_{\rm lt}}{dx_F} +
\frac{d\sigma^S_{\rm iQ}}{dx_F} \, \, .
\label{ismodel}
\ee
The leading-twist distributions are defined in Eqs.~(\ref{fdef}) and 
(\ref{efdef}).  The total intrinsic cross section, $d\sigma^S_{\rm iQ}/dx_F$, 
is related to $dP_S^h/dx_F$ by
\be
\frac{d\sigma^S_{\rm iQ}}{dx_F} = \sigma_{h N}^{\rm in}
\frac{\mu^2}{4 \widehat{m}_s^2} \frac{dP_S^h}{dx_F} \, \, .
\label{iscross}
\ee

We assume that the relative $A$
dependence for leading-twist and intrinsic production is the same as that for
heavy quarks and quarkonia \cite{VB,GutVogt1,VBH2,VBH1}.  The $A$ dependence of
the two component model is
\be
\sigma_{hA} = A^\beta \sigma_{\rm lt} + A^\gamma \sigma_{\rm iQ}
\label{sigadep}
\ee
where the combination of the two terms should approximate $A^\alpha$
in Eq.~(\ref{alfdef}).
There are no strong nuclear effects on open charm at leading twist so that the 
$A$ dependence is linear at $x_F \sim 0$, $\alpha = 1$, \cite{e789d}, 
dropping to $\alpha = 0.77$ for 
pions and 0.71 for protons \cite{VBH2,Badier} at higher $x_F$ where the
intrinsic model begins to dominate. We assume that $\beta = 1$ and $\gamma = 
0.77$ for pions and 0.71 for all baryons.  Thus, 
\be
A^{\gamma - \beta} \approx A^{-1/3} \,\,\,\, {\rm as} \,\,\,\,
x_F \rightarrow 1 \, \,
. \label{tcadep}
\ee
This relative $A$ dependence, similar to that discussed earlier for 
light hadrons \cite{Geist}, is
included in our calculations.  
The proton and neutron numbers are taken into account in the calculation of
the leading-twist cross section.  This isospin effect is small for fusion, $F$,
which is dominated by the $gg$ channel.  In perturbative QCD, $\beta = 1$ 
could be modified by changes in the nuclear parton distributions relative to
the proton \cite{Arneodo}.  However, the scale for our perturbative
calculation is too low for such models of
these modifications to apply \cite{EKS981,EKS982} and are not
included in our calculations.

\section{Results}

\begin{figure}[htpb]
\setlength{\epsfxsize=0.95\textwidth}
\setlength{\epsfysize=0.3\textheight}
\centerline{\epsffile{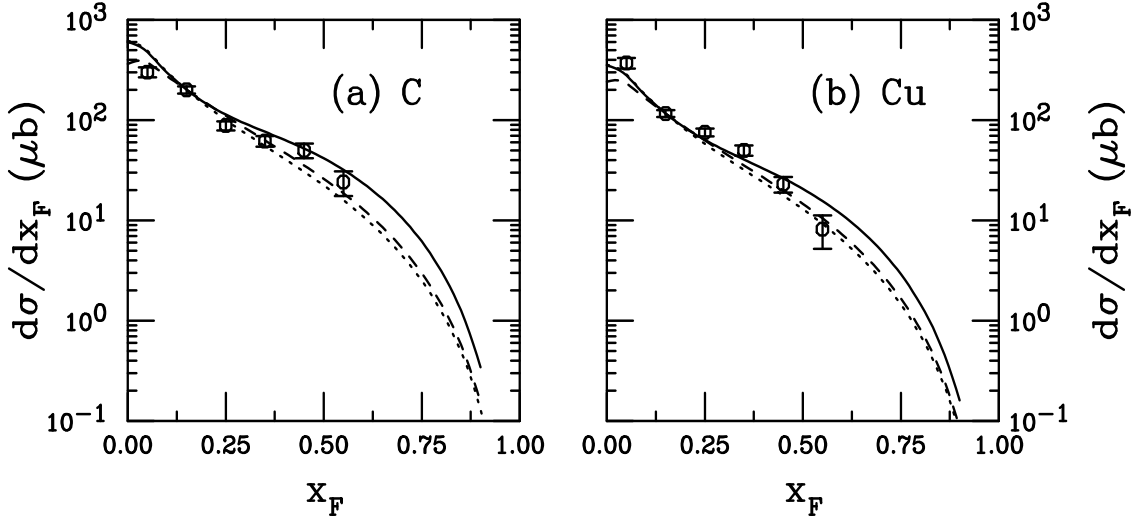}}
\caption[]{The model is compared to the 345 GeV WA89 pion data on
(a) C and (b)
Cu targets.  The dotted curves are leading-twist fusion, $F$, alone, the dashed
curves include uncorrelated fragmentation and coalescence, and the solid
curves include coalescence alone.  The data sets have been normalized to 
the cross section per nucleon.  The curves are normalized to the data at $x_F =
0.15$.}
\label{pipa}
\end{figure}

We begin by comparing the model to the WA89 pion data in Fig.~\ref{pipa}.
These data do not strongly distinguish between leading-twist
fusion and the full model.  The intrinsic results do
not significantly depend on uncorrelated
fragmentation.  All three curves agree rather well with the data, primarily
because the fusion $x_F$ distribution is already fairly hard.  Then the
intrinsic contribution is a small effect even though
the $d$ valence quark is common between the $\pi^-$ and the $\Xi^-$.  This is
perhaps due to the fact that $P_{\rm iss}^6$ is already rather small, $\sim
0.6$\%. We note that the calculated total cross sections agree with the
measured cross sections to within better than a factor of two despite the
rather large uncertainties in the calculations.

Even though the intrinsic contribution is relatively small, it significantly
affects the value of $a$ obtained from Eq.~(\ref{avea}).  The difference
between the $a$ values found without and with uncorrelated fragmentation in the
intrinsic model is
negligible for the pion beam.  We find $a \approx 4.1$ for the C target
and 4.3 for the Cu target relative to $a=5.2$ for leading twist alone.  
These results are within the errors of the WA89 fit
to their data.  The agreement is especially good  
since the two-component model does not give a smooth falloff as a
function of $x_F$ that can be easily quantified by a single exponent.

It is also possible to calculate $\alpha(x_F)$ and the $x_F$-integrated
$\alpha$ from the ratios of the distributions.  The calculations including
both uncorrelated fragmentation and coalescence generally give a smaller value
of $\alpha$ and, hence, a stronger $A$ dependence.  This is because
fragmentation peaks at low $x_F$, influencing the $A$ dependence sooner than
coalescence alone which is only significant at intermediate $x_F$.
Thus $\alpha(x_F) \sim 0.9$ for fragmentation and coalescence while $\alpha$
decreases from $\sim 1$ at low $x_F$ to 0.86 at high $x_F$ with coalescence
alone.  The integrated values are 0.93 and 0.98 respectively, somewhat higher
than the WA89 result but with the same general trend.

\begin{figure}[htpb]
\setlength{\epsfxsize=0.95\textwidth}
\setlength{\epsfysize=0.3\textheight}
\centerline{\epsffile{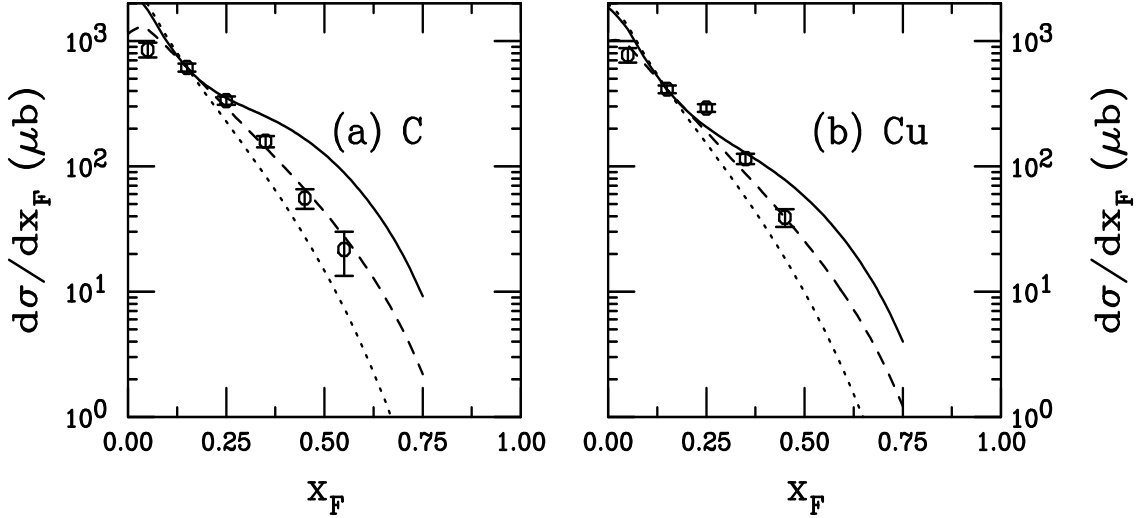}}
\caption[]{The model is compared to the 260 GeV WA89 neutron data on
(a) C and (b)
Cu targets.  The dotted curves are leading-twist fusion, $F$, alone, the dashed
curves include uncorrelated fragmentation and coalescence, and the solid
curves include coalescence alone.  The data sets have been normalized to
the cross section per nucleon. The curves are normalized to the data at $x_F =
0.15$.} 
\label{npa}
\end{figure}

The overall agreement with the total cross section is not as good for the $nA$
data, shown in Fig.~\ref{npa}.  Surprisingly, the distribution
including uncorrelated fragmentation agrees best with
the data.  This is perhaps because the energy of the secondary
neutron beam is least well determined.  The
energy spread is 15\% compared to 9\% for the pion and $\Sigma^-$ beams.  
A small energy variation can have a large effect on the 
leading-twist
cross section.  A 15\% increase in the average neutron energy, from 260 GeV to 
300 GeV, increases $d\sigma_{\rm lt}/dx_F$ by 80\% at $x_F \sim 0.25$ while the
intrinsic cross section is essentially unaffected.  Such a shift in the
relative leading-twist and intrinsic production rates could easily reduce the
effect of coalescence alone to be more compatible with the data.
The uncertainty in the energy of the pion beam has a much weaker effect on the
final result because the intrinsic contribution is already small, as is obvious
from Fig.~\ref{pipa}.

The calculated exponents $a$ are larger for the neutron than the pion, in
agreement with the WA89 measurements \cite{wa89xi}. We find $a \approx 4.9$
for C and 5.8 for Cu.  Typically the value of $a$ obtained for fragmentation
and coalescence is larger than that for coalescence alone since eliminating the
fragmentation contribution tends to increase $\langle x_F \rangle$.  The
stronger $A$ dependence assumed for the intrinsic model has the effect of
increasing $a$ for larger nuclei.  Thus the $a$ found for the carbon target
agrees rather well with the WA89 data while the copper data suggest a harder
distribution than our calculation implies.  There is, however, a stronger 
$A$ dependence in the falloff with $x_F$ than in the data which cannot
distinguish between the values of $a$ determined for the two targets.  This
stronger dependence is reflected in the values of $\alpha$ obtained, 0.87 when
fragmentation and coalescence are included and 0.95 with coalescence alone.

Even though there is some $A$ dependence in the model calculations,
the relatively small intrinsic contribution to the pion and neutron data
leads to a rather weak overall $A$ dependence.  Dominance of the leading-twist
cross section at low to intermediate $x_F$ results in a nearer-to-linear 
integrated $\alpha$, as observed by WA89 \cite{wa89xi}.

We now turn to $\Xi^-$ production by the $\Sigma^-$ where the $A$ dependence
can be expected to be stronger.  For the
first time, we have a valence strange quark in the projectile so that we can
compare the effectiveness of fusion alone with flavor
excitation.  We can also better test the importance of uncorrelated
fragmentation because coalescence production is already possible
in the 5-parton Fock state, $|dds s \overline s
\rangle$. 

\begin{figure}[htpb]
\setlength{\epsfxsize=0.95\textwidth}
\setlength{\epsfysize=0.5\textheight}
\centerline{\epsffile{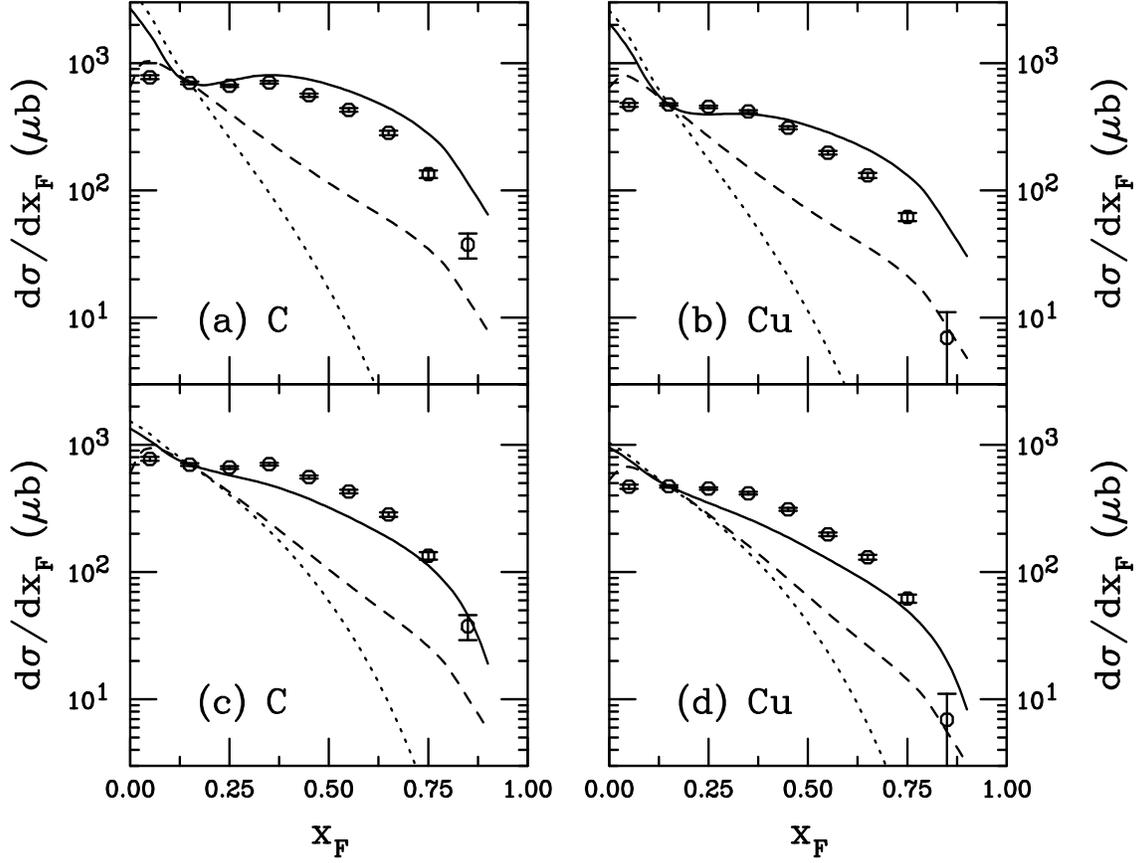}}
\caption[]{The model is compared to the 345 GeV WA89 $\Sigma^- A$ data on C and
Cu targets.  In (a) and (b), the leading twist contribution is $F$ 
while in (c) and (d), flavor excitation is also included, $F +
X_{p_{T_{\rm min}}}$. 
The dotted curves are for leading-twist alone, the dashed
curves include uncorrelated fragmentation and coalescence and the solid
curves include coalescence alone.  The data sets have been normalized to
the cross section per nucleon. The curves are normalized to the data at $x_F =
0.15$.}
\label{sigmpa}
\end{figure}

Our results are collected in Fig.~\ref{sigmpa}.  
We first discuss the
importance of uncorrelated fragmentation to leading-twist
fusion, $F$, alone, Fig.~\ref{sigmpa}(a) and (b).  
The leading-twist contribution is rather steeply falling.
Including both uncorrelated fragmentation and
coalescence broadens the $x_F$ distribution but cannot match the hardness of
the measured $x_F$ distribution.  Eliminating the fragmentation contribution
produces a much harder distribution for $x_F \geq 0.15$, matching the
shape of the data relatively well.

The calculations are all normalized to the $x_F = 0.15$ point to better
compare the shapes of the distributions. Fragmentation gives better agreement 
at low $x_F$ because uncorrelated
fragmentation peaks close to $x_F \sim 0$, filling
in the low to intermediate $x_F$ range.  Coalescence, on the other hand, always
produces strange hadrons with $\langle x_F \rangle \geq 0.3$, broadening the
distribution only in this region.  Thus without 
fragmentation the
calculations overestimate the
data at $x_F \sim 0$.  The data seem to indicate that
uncorrelated fragmentation is not an effective mechanism for intrinsic
production, in agreement with the conclusions of Ref.~\cite{GutVogtIS}.

The agreement with the solid curves in Fig.~\ref{sigmpa}(a) and (b)
is good but not perfect.  The calculation overestimates the data at high $x_F$.
Recall that for the neutron, the 15\% spread in the beam momentum
could result in an overestimate of the intrinsic contribution, as previously
discussed.  Although the possible spread in the $\Sigma^-$ beam momentum
is smaller, it could affect the relative intrinsic 
contribution at low to intermediate 
$x_F$.  At large $x_F$, the effect on the shape would be negligible because the
intrinsic contribution dominates.  Thus, given the inherent uncertainties
in the model and in the data, the agreement is rather satisfactory.

We have obtained the value of the exponent $a$ from $\langle x_F \rangle$,
both over all $x_F$ and for $x_F > 0.1$, avoiding the strong change in slope 
of the solid curves when coalescence alone is included in the intrinsic result.
When the entire forward $x_F$ range is integrated over, $a = 3.02$ for C and
3.39 for Cu with both uncorrelated fragmentation and coalescence while with
coalescence alone, $a = 1.24$ for C and 1.78 for Cu.  Considering only the
range $x_F > 0.1$, we find $a = 1.56$ for C and 1.63 for Cu with fragmentation
and 0.43 for C, 0.57 for Cu without fragmentation.  The calculated $a$'s
suggest considerably harder $x_F$ distributions in the more limited $x_F$
region, particularly when coalescence alone is considered.  However, none of 
the results are in good agreement with $a \approx 2$, as obtained by WA89 for 
$x_F \geq 0.4$.  This is not surprising, especially since the solid curve is
seen to be harder than the data for $x_F > 0.1$.  In any case, the coalescence
contributions in particular, now considerably more important than for $\Xi^-$
production by non-strange hadrons, are difficult to fit to a power law since
they approach zero at both $x_F = 0$ and $x_F = 1$ with a peak at intermediate
$x_F$, see the curves in Ref.~\cite{GutVogtIS}.  The various contributions,
all with somewhat different magnitudes due to the relative probabilities, peak
at different values of $x_F$, complicating the situation further.

The calculated values of $\alpha$ give $\alpha \approx 0.8$ for the integrated
cross sections but $\alpha \approx 0.7$ for $x_F > 0.4$, with and without
fragmentation in the intrinsic model, rather consistent with the WA89 result.
However, as a function of $x_F$, coalescence alone is more consistent with the
measurements since $\alpha \approx 1$ at $x_F \approx 0$, decreasing to 0.71
as $x_F \rightarrow 1$, as expected from Eq.~(\ref{sigadep}). 

We now check if our results improve when we include 
flavor excitation, Eq.~(\ref{efdef}), shown in
Fig.~\ref{sigmpa}(c) and (d).  Now the
baseline leading twist distribution, $F + X_{p_{T_{\rm min}}}$, is harder than
with fusion alone, as shown in Fig.~\ref{fus}(c).  
However, although the distribution
is broader, it still drops six orders of magnitude over the entire $x_F$ range
with $p_{T_{\rm min}} = 1$ GeV.  Thus including
flavor excitation cannot describe the data without the intrinsic coalescence
component, as in
Fig.~\ref{sigmpa}(a) and (b). The total cross section is in reasonable
agreement with that measured by WA89. Decreasing $p_{T_{\rm min}}$ further
can harden the distribution but still underestimates the data.
A lower $p_{T_{\rm min}}$ enhances the total cross sections considerably 
so that, if $p_{T_{\rm min}} = 0.25$ GeV, the cross section is overestimated by
several orders of magnitude.  The
intrinsic contribution is then negligible so that decreasing $p_{T_{\rm min}}$
actually degrades the agreement with the data.  Thus there is no clear evidence
for flavor excitation.

The trends in the $A$ dependence are similar when excitation is included
although the values of $a$ obtained are suggestive of harder $x_F$
distributions than with leading-twist fusion alone.  In particular, the
excitation contribution is harder at low $x_F$, see Fig.~\ref{fus}, causing
the change in slope due to the hardening of the intrinsic distribution when
coalescence alone is included to be less abrupt.  Nonetheless, the agreement
with the measured value of $a$ is not significantly improved.  The calculated
values of $\alpha(x_F)$ are similar to those with leading-twist fusion alone
but the integrated values of $\alpha$ are somewhat larger, $\approx 0.87$,
due to the larger leading-twist baseline.  The $A$ dependence also does not
support flavor excitation as a significant contribution to strange hadron
production. 

To summarize, the $A$ dependence of $\Xi^-$ production by $\Sigma^-$
is stronger because the intrinsic contribution
with coalescence dominates the $x_F$ distribution already at $x_F \sim
0.1$.  Therefore the integrated $A$ dependence is nearly a factor of
$A^{1/3}$ down relative to the pion and neutron data, as shown in
Eqs.~(\ref{sigadep}) and (\ref{tcadep}).  Thus the trends of the model are
in agreement with the WA89 data.

\begin{figure}[htpb]
\setlength{\epsfxsize=0.95\textwidth}
\setlength{\epsfysize=0.5\textheight}
\centerline{\epsffile{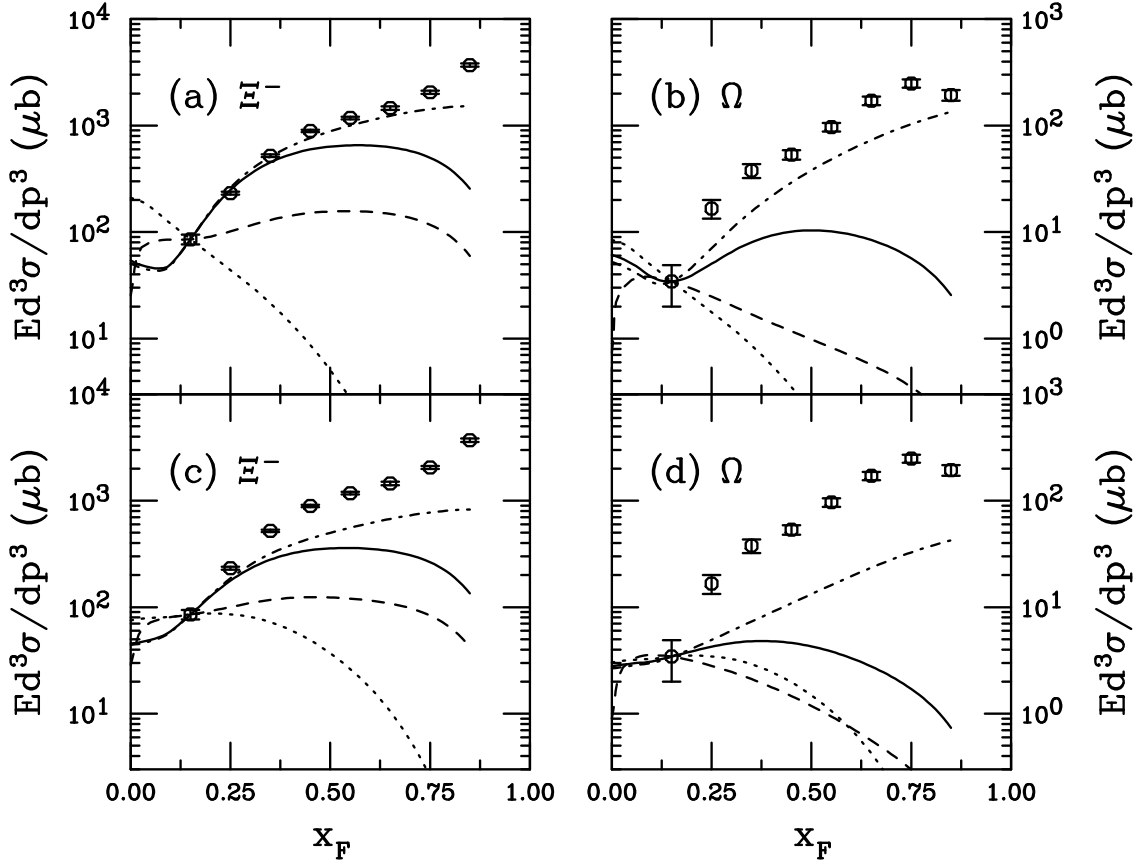}}
\caption[]{The model is compared to the 116 GeV $\Xi^-$Be data.  
In (a) and (b), the leading twist contribution is $F$ 
while in (c) and (d), flavor excitation is also included, $F +
X_{p_{T_{\rm min}}}$. 
The dotted curves are for leading-twist fusion alone, the dashed
curves include uncorrelated fragmentation and coalescence, the solid
curves include coalescence alone and the dot-dashed curves include a
diffractive ``Pomeron'' contribution.  The data sets have been normalized to
the cross section per nucleon. The curves are normalized to the data at $x_F =
0.15$.}  
\label{ximpa}
\end{figure}

Finally, we compare our intrinsic model calculations with the invariant $\Xi^-$
and $\Omega$ cross sections measured in $\Xi^-$Be interactions at 116 GeV
\cite{biagi}.
Since intrinsic production is expected to be a primarily low $p_T$ effect
\cite{VB}, we only compare to the low $p_T^2$ bin, $0<p_T^2 < 0.4$
GeV$^2$ \cite{xiomdat}.  The data and our calculations are 
shown in Fig.~\ref{ximpa}.  We have multiplied our
$x_F$ distributions by $2m_T \cosh y/\sqrt{s}$ to obtain the invariant cross
section.  The invariant $x_F$ distributions are harder as a function of
$x_F$. 

Because the initial and final states are identical for $\Xi^-$ production, the
intrinsic contribution increases with $x_F$.  However, even with coalescence
alone, the increase does not continue beyond $x_F \sim 0.4$.  A similar but
weaker effect is seen for the $\Omega$ where there are two strange quarks in
common with the $\Xi^-$.  Therefore we have tried to identify
a mechanism that would increase the cross section
beyond $x_F \sim 0.4$.  One possibility is a
``Pomeron-like'' parton in the Fock state.  Since the Pomeron has
quantum numbers similar to two gluons, it can be exchanged between two 
projectile valence quarks.  A $|dss {\cal P} \rangle$ state, where ${\cal P}$
signifies the ``Pomeron'', would result in $\Xi^-$ states at high $x_F$ while
avoiding the $\delta(1-x_F)$ delta function for the 3-particle Fock state.
A $\Xi^-$ from such a configuration would have a distribution peaking at $x_F
\rightarrow 1$.  It is also possible to imagine a $|dss s
\overline s {\cal P} \rangle$ state from which both
the $\Xi^-$ and $\Omega$ could
be produced.  In this case, the distribution would peak away from $x_F \sim 1$.
We included ``Pomeron'' production from both states assuming that $P_{{\rm
i}{\cal P}}^4 = P_{\rm iq}^5 \sim 5$\% and that  $P_{{\rm
is}{\cal P}}^6 = P_{\rm iqq}^7 \sim 3.5$\%, giving these configurations large
probability.  The results, shown in the dot-dashed curves in Fig.~\ref{ximpa},
agree relatively well with the data, especially for the $\Xi^-$.
The $\Omega$ data are still underestimated but the trend is now in the right
direction. 

Of course, this is a rather artificial solution, especially when 
the initial and final states are not identical.  If it is correct, it
should also be included in $\Sigma^- p
\rightarrow \Xi^- X$, as shown in Fig.~\ref{sigmpa}.  However, we have checked
this and found that the resulting $x_F$ distribution is far too
hard.  Therefore, the practicality of the mechanism is questionable and the
``Pomeron'' results should not be taken too seriously.

Our calculations with flavor
excitation are compared to the data in Fig.~\ref{ximpa}(c) and (d).
The results do not improve, even when the ``Pomeron'' is
included.  Indeed, the results with excitation underestimate the data at 
high $x_F$ more than with fusion alone.
Therefore we conclude that flavor excitation is
not an effective mechanism for strange hadron production at low $p_T$,
in keeping with the interpretation of the excitation
diagrams as NLO contributions to the production cross section, 
as discussed previously.

We have also calculated the exponent $a'$, see Eq.~(\ref{param2}), for these
distributions with $x_F > 0.5$.  The results are negative for all the cases
shown. We find $a' \approx -0.45$ with and without uncorrelated fragmentation
and $-0.5$ with the ``Pomeron''.  These values are rather consistent with
those obtained for the low $p_T^2$ selection of the $\Xi^-$ data.  The
corresponding values for $\Omega$ production are somewhat lower, $\approx
-0.41$, without the ``Pomeron'' but somewhat higher, $\approx -0.54$, with it.

\section{Conclusions}

We have compared our intrinsic calculations to $\Xi^-$
production by $\pi^-$, $n$ and $\Sigma^-$ projectiles and
to $\Xi^-$ and $\Omega$ production by $\Xi^-$ projectiles.  We find good
agreement with the WA89 data for leading-twist fusion
and coalescence.  Flavor excitation 
seems excluded as a significant mechanism of low $p_T$ strange
hadron production.  The apparent difficulties with uncorrelated fragmentation 
seen in Ref.~\cite{GutVogtIS} are confirmed here.  Thus
coalescence production is the most effective mechanism for strange hadron 
production in the intrinsic model.
The leading charm analysis should perhaps be revisited in light of 
this conclusion.

The conclusions that can be reached from the $\Xi^-$-induced interactions 
at 116 GeV are less clear.  It is possible that a ``Pomeron-like'' 
state could exist in the
hadron wavefunction but its applicability to $\Omega$ production is
somewhat doubtful.  Therefore the interpretation of
these data within the
intrinsic model is rather inconclusive.  More standard studies of diffractive
production in $\Xi^- \, {\rm Be} \, \rightarrow \Xi^- X$ should be performed.

Acknowledgments: We thank
P. Hoyer for reminding us of these data.  R.V. would like to thank
the Niels Bohr Institute and the Grand Accelerateur National d'Ions Lourds
for hospitality at the beginning of this work.

\setcounter{equation}{0}
\renewcommand{\theequation}{\arabic{equation}}
\begin{center}
{\bf Appendix}
\end{center}
\vspace{0.2in}

Here we give the relevant probability distributions in the intrinsic model
for $\Xi^-$ and $\Omega$
production used in our calculations.  To more clearly distinguish between the
probability distributions including uncorrelated fragmentation and
coalescence and those with coalescence alone, both distributions are given.

First, we give the distributions relevant to the WA89 measurements.
We reproduce the $\Xi^-$ probability distribution from a $\pi^-$
\cite{GutVogtIS},
\be
\frac{dP_{\Xi^-}^{\pi^-}}{dx_F} & = & \frac{1}{10} \frac{dP_{\rm
is}^{4F}}{dx_F} + \frac{1}{10} \frac{dP_{\rm
isu}^{6F}}{dx_F} + \frac{1}{10} \frac{dP_{\rm
isd}^{6F}}{dx_F} + 
 \frac{1}{2} \left( \frac{1}{10} \frac{dP_{\rm
iss}^{6F}}{dx_F} + \frac{1}{7} \frac{dP_{\rm iss}^{6C}}{dx_F} \right) \nonumber
\\ &   & + \, \frac{1}{10} \frac{dP_{\rm
isuu}^{8F}}{dx_F} + \frac{1}{10} \frac{dP_{\rm
isud}^{8F}}{dx_F} + \frac{1}{10} \frac{dP_{\rm
isdd}^{8F}}{dx_F} + 
 \frac{1}{2} \left( \frac{1}{10} \frac{dP_{\rm
issu}^{8F}}{dx_F} + \frac{1}{12} \frac{dP_{\rm issu}^{8C}}{dx_F} \right) 
\nonumber \\ &   & + \, \frac{1}{2} \left( \frac{1}{10} \frac{dP_{\rm
issd}^{8F}}{dx_F} + \frac{2}{12} \frac{dP_{\rm issd}^{8C}}{dx_F} \right) +
 \frac{1}{2} \left( \frac{1}{10} \frac{dP_{\rm
isss}^{8F}}{dx_F} + \frac{3}{16} \frac{dP_{\rm isss}^{8C}}{dx_F} \right)\,\, , 
\label{piprobxim} \\
\frac{dP_{\Xi^-}^{\pi^-}}{dx_F} & = & \frac{1}{7} 
\frac{dP_{\rm iss}^{6C}}{dx_F} + \frac{1}{12} 
\frac{dP_{\rm issu}^{8C}}{dx_F} + \frac{2}{12} \frac{dP_{\rm issd}^{8C}}{dx_F}
+ \frac{3}{16} \frac{dP_{\rm isss}^{8C}}{dx_F} \, \, .
\label{piprobximco} 
\ee
From a neutron projectile,
\be
\frac{dP_{\Xi^-}^n}{dx_F} & = & \frac{1}{10} \frac{dP_{\rm
is}^{5F}}{dx_F} + \frac{1}{10} \frac{dP_{\rm
isu}^{7F}}{dx_F} + \frac{1}{10} \frac{dP_{\rm
isd}^{7F}}{dx_F} + \frac{1}{2} \left( \frac{1}{10} \frac{dP_{\rm
iss}^{7F}}{dx_F} + \frac{1}{13} 
\frac{dP_{\rm iss}^{7C}}{dx_F} \right) \nonumber
\\ &   & + \, \frac{1}{10} \frac{dP_{\rm
isuu}^{9F}}{dx_F} + \frac{1}{10} \frac{dP_{\rm
isud}^{9F}}{dx_F} + \frac{1}{10} \frac{dP_{\rm
isdd}^{9F}}{dx_F} +
 \frac{1}{2} \left( \frac{1}{10} \frac{dP_{\rm
issu}^{9F}}{dx_F} + \frac{2}{22} \frac{dP_{\rm issu}^{9C}}{dx_F} \right) 
\nonumber \\ &   & + \, \frac{1}{2} \left( \frac{1}{10} \frac{dP_{\rm
issd}^{9F}}{dx_F} + \frac{3}{22} \frac{dP_{\rm issd}^{9C}}{dx_F} \right) +
 \frac{1}{2} \left( \frac{1}{10} \frac{dP_{\rm
isss}^{9F}}{dx_F} + \frac{6}{28} \frac{dP_{\rm isss}^{9C}}{dx_F} \right)\,\, , 
\label{nprobxim} \\
\frac{dP_{\Xi^-}^n}{dx_F} & = & \frac{1}{13} 
\frac{dP_{\rm iss}^{7C}}{dx_F} + \frac{2}{22} \frac{dP_{\rm issu}^{9C}}{dx_F} 
+ \frac{3}{22} \frac{dP_{\rm issd}^{9C}}{dx_F} + \frac{6}{28} 
\frac{dP_{\rm isss}^{9C}}{dx_F} \, \, . 
\label{nprobximco} 
\ee
The $\Xi^-$ distribution from a $\Sigma^-$ projectile is,
\be
\frac{dP_{\Xi^-}^{\Sigma^-}}{dx_F} & = & \frac{1}{10} \frac{dP_{\rm
iu}^{5F}}{dx_F} + \frac{1}{10} \frac{dP_{\rm
id}^{5F}}{dx_F} + \frac{1}{2} \left( \frac{1}{10} \frac{dP_{\rm
is}^{5F}}{dx_F} + \frac{2}{6} \frac{dP_{\rm is}^{5C}}{dx_F} \right) + 
\frac{1}{10} \frac{dP_{\rm iuu}^{7F}}{dx_F} 
+ \frac{1}{10} \frac{dP_{\rm iud}^{7F}}{dx_F}  
\nonumber \\ &  & + \,
\frac{1}{10} \frac{dP_{\rm idd}^{7F}}{dx_F} 
+ \frac{1}{2} \left( \frac{1}{10} \frac{dP_{\rm
isu}^{7F}}{dx_F} + \frac{1}{13} 
\frac{dP_{\rm isu}^{7C}}{dx_F} \right) 
+ \frac{1}{2} \left( \frac{1}{10} \frac{dP_{\rm
isd}^{7F}}{dx_F} + \frac{3}{13} \frac{dP_{\rm isd}^{7C}}{dx_F} \right) 
\nonumber \\ &   & + \, \frac{1}{2} \left( \frac{1}{10} \frac{dP_{\rm
iss}^{7F}}{dx_F} 
+ \frac{6}{16} \frac{dP_{\rm iss}^{7C}}{dx_F} \right) 
+ \frac{1}{10} \frac{dP_{\rm
iuuu}^{9F}}{dx_F} + \frac{1}{10} \frac{dP_{\rm iuud}^{9F}}{dx_F} 
+ \frac{1}{10} \frac{dP_{\rm
iudd}^{9F}}{dx_F} + \frac{1}{10} \frac{dP_{\rm
iddd}^{9F}}{dx_F} \nonumber \\ &   & + \, 
\frac{1}{2} \left( \frac{1}{10} \frac{dP_{\rm
isuu}^{9F}}{dx_F} + \frac{2}{22} \frac{dP_{\rm isuu}^{9C}}{dx_F} \right) 
+ \frac{1}{2} \left( \frac{1}{10} \frac{dP_{\rm
isud}^{9F}}{dx_F} 
+ \frac{3}{22} \frac{dP_{\rm isud}^{9C}}{dx_F} \right)
\nonumber \\ &   & + \, 
\frac{1}{2} \left( \frac{1}{10} \frac{dP_{\rm
isdd}^{9F}}{dx_F} + \frac{14}{22} \frac{dP_{\rm isdd}^{9C}}{dx_F} \right) +
 \frac{1}{2} \left( \frac{1}{10} \frac{dP_{\rm
issu}^{9F}}{dx_F} + \frac{6}{28} \frac{dP_{\rm issu}^{9C}}{dx_F} \right) 
\nonumber \\ &   & + \, \frac{1}{2} \left( \frac{1}{10} \frac{dP_{\rm
issd}^{9F}}{dx_F} + \frac{9}{28} \frac{dP_{\rm issd}^{9C}}{dx_F} \right) +
 \frac{1}{2} \left( \frac{1}{10} \frac{dP_{\rm
isss}^{9F}}{dx_F} + \frac{12}{32} \frac{dP_{\rm isss}^{9C}}{dx_F} 
\right)\,\, , 
\label{sigprobxim} \\
\frac{dP_{\Xi^-}^{\Sigma^-}}{dx_F} & = & \frac{2}{6} 
\frac{dP_{\rm is}^{5C}}{dx_F} + \frac{1}{13} 
\frac{dP_{\rm isu}^{7C}}{dx_F} + \frac{3}{13} \frac{dP_{\rm isd}^{7C}}{dx_F} 
+ \frac{6}{16} \frac{dP_{\rm iss}^{7C}}{dx_F} + \frac{2}{22} 
\frac{dP_{\rm isuu}^{9C}}{dx_F} 
\nonumber \\ &  & + \, \frac{3}{22} \frac{dP_{\rm isud}^{9C}}{dx_F}
+ \frac{14}{22} \frac{dP_{\rm isdd}^{9C}}{dx_F} + \frac{6}{28} 
\frac{dP_{\rm issu}^{9C}}{dx_F} + \frac{9}{28} \frac{dP_{\rm issd}^{9C}}{dx_F}
+ \frac{12}{32} \frac{dP_{\rm isss}^{9C}}{dx_F} \, \, .
\label{sigprobximco} 
\ee

We now present the relevant probability distributions for $\Xi^-$ and $\Omega$
production from a $\Xi^-$ projectile.  First we give the $\Xi^-$ distributions,
\be 
\frac{dP_{\Xi^-}^{\Xi^-}}{dx_F} & = & 
\frac{1}{2} \left( \frac{1}{10} \frac{dP_{\rm
iu}^{5F}}{dx_F} + \frac{1}{6} \frac{dP_{\rm iu}^{5C}}{dx_F} \right) 
+ \frac{1}{2} \left( \frac{1}{10} \frac{dP_{\rm
id}^{5F}}{dx_F} + \frac{2}{6} \frac{dP_{\rm id}^{5C}}{dx_F} \right) \nonumber
\\ &  & + \, \frac{1}{2} \left( \frac{1}{10} \frac{dP_{\rm
is}^{5F}}{dx_F} + \frac{3}{7} \frac{dP_{\rm is}^{5C}}{dx_F} \right) + 
 \frac{1}{2} \left( \frac{1}{10} \frac{dP_{\rm
iuu}^{7F}}{dx_F} + \frac{1}{13} \frac{dP_{\rm iuu}^{7C}}{dx_F} \right) 
\nonumber
\\ &   & + \, \frac{1}{2} \left( \frac{1}{10} \frac{dP_{\rm
iud}^{7F}}{dx_F} + \frac{2}{13} \frac{dP_{\rm iud}^{7C}}{dx_F} \right) +
 \frac{1}{2} \left( \frac{1}{10} \frac{dP_{\rm
idd}^{7F}}{dx_F} + \frac{3}{13} \frac{dP_{\rm idd}^{7C}}{dx_F} \right) 
\nonumber \\ &  & + \,
 \frac{1}{2} \left( \frac{1}{10} \frac{dP_{\rm
isu}^{7F}}{dx_F} + \frac{3}{16} 
\frac{dP_{\rm isu}^{7C}}{dx_F} \right) +
\frac{1}{2} \left( \frac{1}{10} \frac{dP_{\rm
isd}^{7F}}{dx_F} + \frac{6}{16} \frac{dP_{\rm isd}^{7C}}{dx_F} \right)
\nonumber \\ &   & + \, 
 \frac{1}{2} \left( \frac{1}{10} \frac{dP_{\rm
iss}^{7F}}{dx_F} 
+ \frac{6}{17} \frac{dP_{\rm iss}^{7C}}{dx_F} \right) +
 \frac{1}{2} \left( \frac{1}{10} \frac{dP_{\rm
iuuu}^{9F}}{dx_F} + \frac{1}{22} \frac{dP_{\rm iuuu}^{9C}}{dx_F} \right)
\nonumber \\ &   & + \,
 \frac{1}{2} \left( \frac{1}{10} \frac{dP_{\rm
iuud}^{9F}}{dx_F} 
+ \frac{2}{22} \frac{dP_{\rm iuud}^{9C}}{dx_F} \right) +
 \frac{1}{2} \left( \frac{1}{10} \frac{dP_{\rm
iudd}^{9F}}{dx_F} + \frac{3}{22} \frac{dP_{\rm iudd}^{9C}}{dx_F} \right)
\nonumber \\ &   & + \, 
 \frac{1}{2} \left( \frac{1}{10} \frac{dP_{\rm
iddd}^{9F}}{dx_F} 
+ \frac{4}{22} \frac{dP_{\rm iddd}^{9C}}{dx_F} \right) +
 \frac{1}{2} \left( \frac{1}{10} \frac{dP_{\rm
isuu}^{9F}}{dx_F} + \frac{3}{28} \frac{dP_{\rm isuu}^{9C}}{dx_F} \right)
\nonumber \\ &   & + \, 
 \frac{1}{2} \left( \frac{1}{10} \frac{dP_{\rm
isud}^{9F}}{dx_F} 
+ \frac{6}{28} \frac{dP_{\rm isud}^{9C}}{dx_F} \right) +
 \frac{1}{2} \left( \frac{1}{10} \frac{dP_{\rm
isdd}^{9F}}{dx_F} + \frac{9}{28} \frac{dP_{\rm isdd}^{9C}}{dx_F} \right)
\nonumber \\ &   & + \, 
 \frac{1}{2} \left( \frac{1}{10} \frac{dP_{\rm
issu}^{9F}}{dx_F} + \frac{6}{31} \frac{dP_{\rm issu}^{9C}}{dx_F} \right) +
 \frac{1}{2} \left( \frac{1}{10} \frac{dP_{\rm
issd}^{9F}}{dx_F} + \frac{12}{31} \frac{dP_{\rm issd}^{9C}}{dx_F} \right)
\nonumber \\ &   & + \, 
 \frac{1}{2} \left( \frac{1}{10} \frac{dP_{\rm
isss}^{9F}}{dx_F} + \frac{12}{37} 
\frac{dP_{\rm isss}^{9C}}{dx_F} \right)\,\, , \\
\label{xiprobxi} 
\frac{dP_{\Xi^-}^{\rm \Xi^-}}{dx_F} & = & 
\frac{1}{6} \frac{dP_{\rm iu}^{5C}}{dx_F} 
+ \frac{2}{6} \frac{dP_{\rm id}^{5C}}{dx_F} + \frac{3}{7} 
\frac{dP_{\rm is}^{5C}}{dx_F} + \frac{1}{13} \frac{dP_{\rm iuu}^{7C}}{dx_F} 
+ \frac{2}{13} \frac{dP_{\rm iud}^{7C}}{dx_F} + \frac{3}{13} 
\frac{dP_{\rm idd}^{7C}}{dx_F} 
\nonumber \\ &  & + \,
\frac{3}{16} 
\frac{dP_{\rm isu}^{7C}}{dx_F} + \frac{6}{16} \frac{dP_{\rm isd}^{7C}}{dx_F} 
+ \frac{6}{17} \frac{dP_{\rm iss}^{7C}}{dx_F} 
+ \frac{1}{22} \frac{dP_{\rm iuuu}^{9C}}{dx_F} 
+ \frac{2}{22} \frac{dP_{\rm iuud}^{9C}}{dx_F} 
+ \frac{3}{22} \frac{dP_{\rm iudd}^{9C}}{dx_F} \nonumber
\\ &   & + \, 
\frac{4}{22} \frac{dP_{\rm iddd}^{9C}}{dx_F} 
+ \frac{3}{28} \frac{dP_{\rm isuu}^{9C}}{dx_F} 
+ \frac{6}{28} \frac{dP_{\rm isud}^{9C}}{dx_F} 
+ \frac{9}{28} \frac{dP_{\rm isdd}^{9C}}{dx_F} \nonumber \\ &   & + \,
\frac{6}{31} \frac{dP_{\rm issu}^{9C}}{dx_F} 
+ \frac{12}{31} \frac{dP_{\rm issd}^{9C}}{dx_F} + \frac{12}{37} 
\frac{dP_{\rm isss}^{9C}}{dx_F} \,\, .
\label{xiprobxico} 
\ee
Finally, we give the $\Omega$ distribution from a $\Xi^-$ projectile,
\be
\frac{dP_{\Omega}^{\rm \Xi^-}}{dx_F} & = & \frac{1}{10} \frac{dP_{\rm
iu}^{5F}}{dx_F} + \frac{1}{10} \frac{dP_{\rm
id}^{5F}}{dx_F} + \frac{1}{2} \left( \frac{1}{10} \frac{dP_{\rm
is}^{5F}}{dx_F} + \frac{1}{7} \frac{dP_{\rm is}^{5C}}{dx_F} \right) + 
\frac{1}{10} \frac{dP_{\rm iuu}^{7F}}{dx_F} 
+ \frac{1}{10} \frac{dP_{\rm iud}^{7F}}{dx_F}  
\nonumber \\ &  & + \,
\frac{1}{10} \frac{dP_{\rm idd}^{7F}}{dx_F} 
+ \frac{1}{2} \left( \frac{1}{10} \frac{dP_{\rm
isu}^{7F}}{dx_F} + \frac{1}{16} 
\frac{dP_{\rm isu}^{7C}}{dx_F} \right) 
+ \frac{1}{2} \left( \frac{1}{10} \frac{dP_{\rm
isd}^{7F}}{dx_F} + \frac{1}{16} \frac{dP_{\rm isd}^{7C}}{dx_F} \right) 
\nonumber \\ &   & + \, \frac{1}{2} \left( \frac{1}{10} \frac{dP_{\rm
iss}^{7F}}{dx_F} 
+ \frac{3}{17} \frac{dP_{\rm iss}^{7C}}{dx_F} \right) 
+ \frac{1}{10} \frac{dP_{\rm
iuuu}^{9F}}{dx_F} + \frac{1}{10} \frac{dP_{\rm iuud}^{9F}}{dx_F} 
+ \frac{1}{10} \frac{dP_{\rm
iudd}^{9F}}{dx_F} + \frac{1}{10} \frac{dP_{\rm
iddd}^{9F}}{dx_F} \nonumber \\ &   & + \, 
\frac{1}{2} \left( \frac{1}{10} \frac{dP_{\rm
isuu}^{9F}}{dx_F} + \frac{1}{28} \frac{dP_{\rm isuu}^{9C}}{dx_F} \right) 
+ \frac{1}{2} \left( \frac{1}{10} \frac{dP_{\rm
isud}^{9F}}{dx_F} 
+ \frac{1}{28} \frac{dP_{\rm isud}^{9C}}{dx_F} \right)
\nonumber \\ &   & + \, 
\frac{1}{2} \left( \frac{1}{10} \frac{dP_{\rm
isdd}^{9F}}{dx_F} + \frac{1}{28} \frac{dP_{\rm isdd}^{9C}}{dx_F} \right) +
 \frac{1}{2} \left( \frac{1}{10} \frac{dP_{\rm
issu}^{9F}}{dx_F} + \frac{3}{31} \frac{dP_{\rm issu}^{9C}}{dx_F} \right) 
\nonumber \\ &   & + \, \frac{1}{2} \left( \frac{1}{10} \frac{dP_{\rm
issd}^{9F}}{dx_F} + \frac{3}{31} \frac{dP_{\rm issd}^{9C}}{dx_F} \right) +
 \frac{1}{2} \left( \frac{1}{10} \frac{dP_{\rm
isss}^{9F}}{dx_F} + \frac{10}{37} \frac{dP_{\rm isss}^{9C}}{dx_F} 
\right)\,\, , 
\label{xiprobom} 
\\
\frac{dP_{\Omega}^{\rm \Xi^-}}{dx_F} & = & 
\frac{1}{7} \frac{dP_{\rm is}^{5C}}{dx_F} 
+ \frac{1}{16} \frac{dP_{\rm isu}^{7C}}{dx_F} 
+ \frac{1}{16} \frac{dP_{\rm isd}^{7C}}{dx_F} 
+ \frac{3}{17} \frac{dP_{\rm iss}^{7C}}{dx_F}
+ \frac{1}{28} \frac{dP_{\rm isuu}^{9C}}{dx_F} 
+ \frac{1}{28} \frac{dP_{\rm isud}^{9C}}{dx_F} 
\nonumber \\ &   & + \, 
\frac{1}{28} \frac{dP_{\rm isdd}^{9C}}{dx_F} 
+ \frac{3}{31} \frac{dP_{\rm issu}^{9C}}{dx_F} 
+ \frac{3}{31} \frac{dP_{\rm issd}^{9C}}{dx_F} 
+ \frac{10}{37} \frac{dP_{\rm isss}^{9C}}{dx_F} 
\,\, .
\label{xiprobomco} 
\ee

\end{document}